\documentclass{xpkas}

\def\year{ } 
\def\month{ } 
\def\volume{00} 
\def\issue{0} 
\def\beginpage{1} 
\def\endpage{99} 
\def\received{2015} 
\def\accepted{2015} 
\date{Received \received ; accepted \accepted}

\title{
The \textit{AKARI} FIS catalogue of YSOs and extragalactic objects
}

\author[1]{L.~Viktor~T\'oth}
\author[2]{G\'abor~Marton}
\author[1,3]{Sarolta~Zahorecz}
\author[1,2]{Lajos~G.~Bal\'azs}
\author[4,1]{Andrea~Nagy}

\affil[1]{Department of Astronomy, Eotvos University Budapest; \email{l.v.toth@astro.elte.hu}}
\affil[2]{Konkoly Observatory, Research Centre for Astronomy and Earth Sciences, Hungarian Academy of Sciences, Budapest; \email{marton.gabor@csfk.mta.hu}}
\affil[3]{ESO, Garching; \email{szahorec@eso.org}}
\affil[4]{Ferenc Rakoczi II. Transcarpathian Hungarian Institute}

\def\corrauthor{
L.V. T\'oth
}

\def\runningauthor{
L.V. T\'oth
}

\def\runningtitle{
\textit{AKARI} FIS
}

\def\keywords{
infrared: galaxies; open clusters and associations: individual: IC348; stars: formation
}

\def\abstracttext{
The point sources in the Bright Source Catalogue of the \textit{AKARI} Far-Infrared Surveyor (FIS) were classified based on their FIR and mid-IR fluxes and clours into young stellar object (YSO) and extragalactic source types using Quadratic Discriminant Analysis method (QDA) and Support Vector Machines (SVM). The reliability of the selection of YSO candidates is high, and the number of known YSO candidates were increased significantly, that we demonstrate in the case of the nearby open cluster {IC348}.
Our results show that we can separate galactic and extragalactic \textit{AKARI} point sources in the multidimensioal space of FIR fluxes and colours with high reliability, however, differentiating among the extragalactic sub-types needs further information.
}

\begin{document}
\pkashead 


\section{Introduction\label{sec:intro}}
The \textit{AKARI} FIS Bright Source Catalogue (BSC, \citet{yamamura2010}) as an all-sky catalogue opens the possibility to increase the number of known infrared bright YSOs and extragalactic objects. There were already successful attempts to separate point source types, eg. stars and galaxies by \citet{pollo2010}, and AGNs by \citet{ichikawa2012}. \citet{wang2014} classified \textit{AKARI} point sources into various extragalactic types using multi wavelength photometry and spectral energy distribution (SED) templates. In our previous paper \citep{toth2014} statistical methods were applied on \textit{AKARI} and \textit{WISE} photometric data to separate stellar, YSO and extragalactic types of AKARI point sources. We showed that the AKARI YSOs are often associated with Planck cold clumps \citep{planck2011b}, and have an overdensity towards galactic shells \citep{toth2014}. While testing the overall distribution of galactic star formation with the AKARI YSO catalogue we also hope to locate and describe YSO clusters (see eg. \cite{toth2013}). In this paper we recalculate the reliability of our YSO classification and demonstrate the use of our catalogue of \textit{AKARI} YSO candidates. We also report the results of our attempt to identify AKARI extragalactic point source types.

\section{Analysis of the \textit{AKARI} FIS point sources \label{sec:2}}
The \textit{AKARI} FIS BSC lists 427071 point sources. In our study we used sources with the highest quality flag ``3'' in the two wide-filter bands ("WIDE-S" and "WIDE-L", with band centres of "90\,$\mu$m and "140\,$\mu$m" respectively). That means confirmed sources with reliable flux densities ($\approx 20$\% uncertainty). We combined \textit{AKARI} and \textit{WISE} \citep{cutri2012} all-sky photometric data (see \citet{toth2014} for further details). 

\subsection{Object types, colours and selection\label{sec:21}}
The first step of the classification was an investigation of point sources with already known types. We used a search radius of 30$^{\prime\prime}$ to find associated entries in the SIMBAD database. The mid-IR and FIR colour distribution of extragalactic objects, YSOs and reddened stars are similar, thus those are located in highly overlapping regions on the colour--colour and brightness--colour planes. For example, the average [F90/F140] colour of the SIMBAD "Y*O" (YSO), "G" (galaxy) and "pA*" (post-AGB star) types are $-0.511\pm0.238$, $-0.513\pm0.24$ and $-0.536\pm0.231$, respectively.  Separating amoprhous regions in the multi-dimensional parameter space can not be made by cutting it with a few planes. We can handle that with advanced statistical methods, like Quadratic Discriminant Analysis. The classification was based on a 6D parameter space that contained the \textit{AKARI} FIS BSC [F65/F90], [F90/F140], [F140/F160] colours and the [F140] flux along with the WISE W1-W2 colour and the W1 magnitude.

\subsection{Quality check of the selection\label{sec:22}}
As it was described in \citet{toth2014} the \textit{AKARI} FIS BSC sources were classified into 3 main objects types, namely candidate YSOs, evolved stars and galaxies. The goodness of our automatic classification was measured comparing the total number of candidates of a given type listed also in SIMBAD to the number of those which had the correct SIMBAD object types. For our YSO candidates the approving SIMBAD types were "Y*O" (YSO), "TT*" (TTauri), "pr*" (pre-main sequence star), "Or*" (variable star of Orion type) and "FU*"(variable star of FU Ori type). The fraction of contaminating sources was then estimated using the number of SIMBAD associates with any other SIMBAD object type.

A refreshed list of the SIMBAD associations was created as compared to \citet{toth2014}, and the estimation of the goodness and contamination was recalculated. In case of the YSO candidates we were able to keep 84.7\% of the YSO-like sources.


We found 18494 sources that had colours and brightness values similar to extragalactic objects, but a large number of SIMBAD stars were identified among them. Therefore we created a training sample tailored for identification of extragalctic sources by using SIMBAD approved galaxies located outside the Galactic mid-plane ($|b|>3^{\circ}$). This way 5138 extragalctic source candidates were identified using the QDA method. 4451 (86.6\%) of them are known extragalactic objects based on their SIMBAD identification. 297 more sources were identified as "IR" infrared source or "Rad" radio source. 262 sources were identified as some kind of Galactic object type and 128 remained unidentified.

\begin{figure}[h]
\centering
\includegraphics[width=80mm]{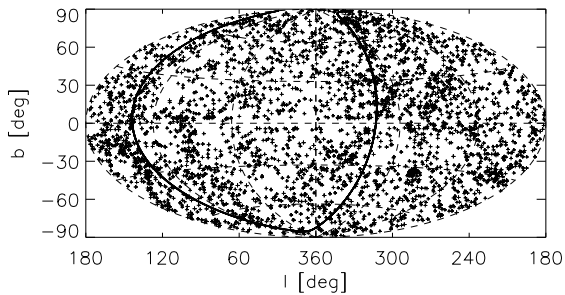}
\caption{Celestial distribution of extragalactic objects classified with QDA in galactic coordinate system, using Aitoff-projection. The super galactic equator is shown with the thick solid black line.\label{fig:fig3}}
\end{figure}

The all-sky distribution of our \textit{AKARI} extragalactic candidate point sources has an overdensity along the supergalactic plane as shown in Figure \ref{fig:fig3}. It may indicate that a considerable number of these objects are members of the Local Supercluster. There are also other local density enhancements seen, an interpretation of those needs further investigations.

\section{Use of the \textit{AKARI} catalogues\label{sec:3}}

\subsection{New YSOs in IC348\label{sec:31}}
IC348 is a 2-3 Myr old cluster, located at the edge of the Perseus molecular cloud. It contains 283 spectroscopically confirmed members in a relatively compact ($20^{\prime\prime}\times 20^{\prime\prime}$) region \citep{lee2011}. It is located at a distance of 320\,pc, based on \citet{herbig1998}. The cloud G159.6-18.5 was recognised as an HII region and a dusty shell of enchanced FIR emission with a diameter of 1.5 deg by \citet{anderson2000} and \citet{watson2005}.

\begin{figure}[h]
\centering
\includegraphics[width=80mm]{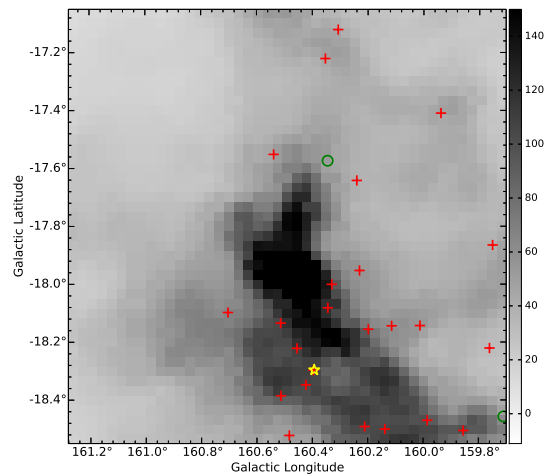}
\caption{New candidate YSOs in IC348. Class I (red plus) and Class II (green circle) YSOs are overlaid on the Planck 857 GHz image. The 857GHz (350$\mu$m) intensity varies from 10MJysr$^{-1}$ to over 140MJysr$^{-1}$ in the image \citep{planck2011} showing IC348 and the G159.6-18.5 HII region. The yellow star marks the position of the Class I source AKARI-FIS-V1 J0342443+315028. See also its SED in Figure \ref{fig:fig1}). \label{fig:fig0}}
\vspace{5mm} 
\end{figure}

We found 26 new YSO candidates in the IC348 cluster. The distribution of these \textit{AKARI} FIS YSOs is shown in, Figure\ref{fig:fig0}. As part of our survey of the Taurus-Auriga-Perseus region \citep{toth2015} young stellar objects were classified into different evolutionary classes, based on their spectral index, $\alpha$ or bolometric temperature, $T_{bol}$. The $\alpha$ index is the slope of the near- and mid-infrared part of the spectral energy distribution (see \citet{lada1987}, \citet{greene1994} and \citet{andre1993}), while $T_{bol}$ is defined as the temperature of a blackbody, whose spectrum has the same mean frequency as the observed spectrum \citep{myers1993}. 
Class 0 objects are undetectable at $\lambda<20\mu$m, their bolometric temperature is below 70 K. The spectral index ranges for Class I; Flat spectrum; Class II; and Class III YSO types are: $\alpha>0.3$; $0.3>\alpha>-0.3$; $-0.3> \alpha>-1.6$; and $-1.6>\alpha$ respectively. The bolometric temperature ranges for Class I; Class II; and Class III source types are: $70$K$<T_{bol}<650$K; $650$K$<T_{bol}<2800$K; and $T_{bol}>2800$K respectively.

There are 24 Class I and 2 Class II sources among the newly discovered YSOs of IC348, as listed in Table \ref{table:tab0}. The YSOs are mostly concentrated at the dusty ISM that is seen as a bright extended region in the Planck 857 GHz image (Figure \ref{fig:fig0}). The Lada classes are marked with red pluses (Class I) and green circles (Class II).

\begin{table}[!hpbt]
\footnotesize\addtolength{\tabcolsep}{-5pt}
\begin{tabular}{c | r r | c}
\textit{AKARI} name & L [deg] & B [deg] & Lada class \\
\hline
0339531+320754	&	159.710	&	-18.457	&	Class II	\\
0341483+323334	&	159.751	&	-17.865	&	Class I	\\
0340454+321703	&	159.761	&	-18.221	&	Class I	\\
0340146+315958	&	159.855	&	-18.506	&	Class I	\\
0343508+324756	&	159.938	&	-17.410	&	Class I	\\
0340484+315703	&	159.985	&	-18.471	&	Class I	\\
0341526+321102	&	160.012	&	-18.144	&	Class I	\\
0342119+320746	&	160.114	&	-18.145	&	Class I	\\
0341140+314950	&	160.138	&	-18.502	&	Class I	\\
0342291+320359	&	160.198	&	-18.157	&	Class I	\\
0341327+314811	&	160.211	&	-18.493	&	Class I	\\
0343123+321211	&	160.230	&	-17.954	&	Class I	\\
0344111+322613	&	160.240	&	-17.643	&	Class I	\\
0346006+324743	&	160.307	&	-17.122	&	Class I	\\
0343259+320626	&	160.330	&	-18.002	&	Class I	\\
0343125+320209	&	160.344	&	-18.083	&	Class I	\\
0344465+322554	&	160.345	&	-17.575	&	Class II	\\
0345514+324116	&	160.353	&	-17.222	&	Class I	\\
0342443+315028	&	160.396	&	-18.299	&	Class I	\\
0342402+314714	&	160.423	&	-18.349	&	Class I	\\
0343095+315136	&	160.455	&	-18.223	&	Class I	\\
0342216+313650	&	160.483	&	-18.524	&	Class I	\\
0342531+314200	&	160.513	&	-18.387	&	Class I	\\
0343395+315339	&	160.514	&	-18.136	&	Class I	\\
0345292+321931	&	160.539	&	-17.553	&	Class I	\\
0344254+314817	&	160.704	&	-18.099	&	Class I	\\
\end{tabular}
\caption{New candidate YSOs towards the IC348 open cluster. The columns are: \textit{AKARI} name; galactic coordinates L and B; and YSO class \citep{lada1987}.\label{table:tab0}}
\end{table}

The stellar parameters (e.g. stellar mass, age, temperature; disk and envelope mass and size) of the candidate YSOs were estimated using online available pre-computed YSO SEDs\footnote{http://caravan.astro.wisc.edu/protostars/}. \citet{robitaille2007} presented a grid of 20000 radiation transfer models of YSOs in different evolutionary stages and from ten different viewing angles, resulting in 200000 SEDs. The best fitting Robitaile models provided us the parameter values of AKARI YSO candidates. To estimate the error of the fit the 9 next best fitting models were used, the minima and maxima of each of the parameters were derived from those. Our succesful SED fits also support the validity of our statistical YSO selection.

\begin{figure}[h]
\centering
\includegraphics[width=80mm]{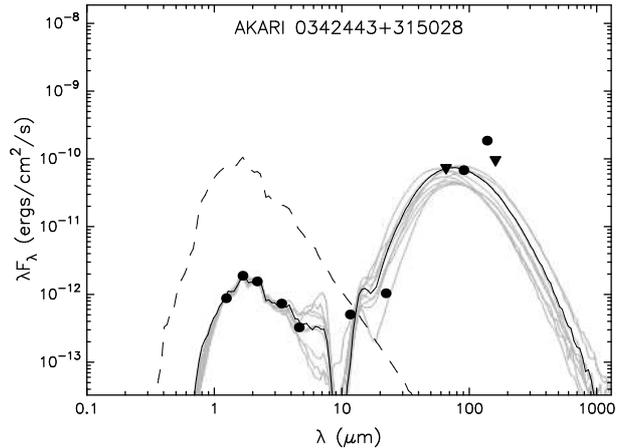}
\caption{SED of AKARI-FIS-V1 J0342443+315028, a YSO in IC348, fitted with the YSO models of \citet{robitaille2007}. Filled circles show the fitted 2MASS, \textit{WISE} and \textit{AKARI} fluxes, triangles indicate upper limits. The solid black line shows the best-fitting model, and the grey lines show the 9 next best models. The dashed line shows the stellar photosphere of the best-fitting model. \label{fig:fig1}}
\end{figure}

A sample of the fitted SEDs is shown in Figure \ref{fig:fig1} for AKARI-FIS-V1 J0342443+315028, a new YSO without any former SIMBAD association.
We classified the source as a Class I / Flat object based on its $T_{bol}=643$K and $\alpha=-0.1$ values.
The parameters from the 10 best fits (with minimum, maximum values in parentheses) are the following: age is $t\approx 2.1\times10^5$yr ($5.8\times10^4$yr$ < t < 2.3 \times10^5$yr), stellar mass is $M\approx 0.30$M$_\odot$ (0.15M$_\odot < M <0.35$M$_\odot$) , stellar radius is $r\approx 2.6$R$_\odot$ (2.4R$_\odot <R<3.2$R$_\odot$) R$_\odot$, temperature is $T\approx 3350$K (2900K$<T<3500$K) and disk mass is $M_{disk}\approx 1.6\times10^{-3}$M$_\odot$ (5.6$\times10^{-5}$M$_\odot<M_{disk}< 1.3\times10^{-2}$ M$_\odot$).

\subsection{New extragalactic objects?\label{sec:32}}

We searched counterparts of the 128 extragalactic candidates without SIMBAD identification in the Rowan-Robinson catalogue of galaxies \citep{wang2014} using a 60$^{\prime\prime}$ search radius, but did not find any. We note that only 2 of these sources have an associated \textit{IRAS} point source. A crosscorrelation with the NASA/IPAC Extragalactic Database (NED) shows, that 30\% of our 128 SIMBAD unidentified point sources are galaxies, few of them are part of a galaxy, but most of them are either not listed in the NED, or with an object type of radio or infrared source. We considered these as new extragalactic object candidates (NEOCs). Checking the optical and infrared appearance of the NEOCs individually is underway, detailed results will be published elsewhere.

\subsection{Classification into extragalactic subtypes}

We cross-correlated our QDA galaxy candidates with the Rowan-Robinson catalogue \citep{wang2014}, and created a training sample based on their classifications. They listed 7 sub-types for their identified objects: (1) cirrus dominated, (2) M82 dominated, (3) A220 dominated, (4) AGN dust torus dominated, at 60 $\mu$m, (5) ysb dominated, (6) cool cirrus dominated, (7) detected at 60 $\mu$m only and used subtype (0) for point sources where no template contributes over 50\%. 

In order to achieve the best possible result we carried out various classifications with different statistical approaches. Linear and Quadratic Discriminant Analysis (LDA, QDA) are described in more details in our previous paper \citep{toth2014}. Support Vector Machines (SVMs) are a class of supervised learning algorithms, created as an extension to nonlinear models of the generalized portrait algorithm developed by Vladimir Vapnik \citep{vapnik1995} for classification in a multidimensional parameter space. A detailed description can be found in \citet{malek2013}. Our results of the various methods are summarised in Table \ref{table1}, that shows that we can separate cirrus dominated objects (type 1 of \citet{wang2014}) from the other sub-types, but we cannot achieve the same level of confidence with the other sub-types with any of our statistical methods.

\begin{table*}[!hpbt]
\footnotesize\addtolength{\tabcolsep}{-5pt}
\begin{tabular}{c|c||c|c|c|c|c|c|c|c|c|c|c}

Object type&known&LDA&QDA&SVM1&SVM2&SVM3&SVM4&SVM5&SVM6&SVM7&SVM8& SVM9\\
\hline
\hline
no template contributes $>$ 50\% &5&0.0 &0.0 &0.0 &0.0 &20.0 &0.0 &0.0 &0.0 &0.0 &0.0 &0.0\\
\hline
cirrus dominated &2039&96.1 &88.3 &91 &93.7 &98.6 &97.4 &97.7 &98.7 &89.5 &67.4 &66.4\\
\hline
M82 dominated &203&3.4 &20.2 &41.9 &43.4 &19.7 &2.0 &0 &4.9 &45.8 &76.8 &73.9\\
\hline
A220 dominated &502&29.8 &37.6 &22.3 &0.4 &2.2 &38.2 &33.1 &22.1 &20.9 &3.8 &1.8\\
\hline
AGN dust torus dominated, at 60 $\mu$m &10&50.0 &0.0 &10.0 &0.0 &10.0 &0.0 &0.0 &20.0 &10.0 &0.0 &10.0\\
\hline
ysb dominated &101&32.7 &34.7 &9.9 &0 &3.9 &23.8 &0 &24.8 &8.9 &0 &6.9\\
\hline
cool cirrus dominated &111&3.6 &33.3 &0 &0 &0.9 &0 &0 &0 &0 &0 &0.9\\
\hline
Average &&30.8 &30.6 &25.0 &19.6 &22.2 &23.1 &18.7 &24.4 &25.0 &21.1 &22.8\\
\end{tabular}
\caption{\citet{wang2014} object types and the reliability of classification into each types using various classification algorithms. The reliability is measured as the ratio of sources which were reclassified into their own group by our selection method. SVM types from 1-9 correspond to the eps-regression type with radial, linear and polynomial kernel, the C-Classification method with radial, linear and polynomial kernel and to the nu-regression method with radial, linear and polynomial kernel, again.\label{table1}
}
\end{table*}

\section{Conclusions}
Our results show that we can distinguish YSO and extragalactic source candidates from the other types of objects with high reliability applying QDA method for object classification on \textit{AKARI} FIS BSC and \textit{WISE} data. We also conclude that differentiating among the extragalactic sub-types needs further information. Follow-ups may approve our classification of the NEOCs.



\acknowledgments
Valuable comments by Michael Rowan-Robinson and Glenn White and our anonymous referee are acknowledged. This research has made use of the NASA/IPAC Extragalactic Database (NED) which is operated by the Jet Propulsion Laboratory, California Institute of Technology, under contract with the National Aeronautics and Space Administration. This research was supported by the OTKA grants NN 111016 and K101393.




\end{document}